\documentstyle[11pt,aaspp4,flushrt]{article}  %full text width preprint style 
\lefthead{Raig, Gonz\'alez Casado \& Salvador-Sol\'e}
\righthead{Scaling Evolution of Halo Density Profile}

\def\roc{\rho_{\rm c}}
\def\rs{r_{\rm s}}
\def\delc{\delta_{\rm c}}
\def\xs{x_{\rm s}}
\def\xsi{x_{{\rm s}i}}

\def\xsf{x_{\rm sf}}
\def\tf{t_{\rm f}}
\newcommand{\der}{{\rm d}}
\def\delm{\Delta_{\rm m}}

\begin{document}

\title{SCALING EVOLUTION OF UNIVERSAL DARK-MATTER HALO DENSITY PROFILES}

\author{Andreu Raig$^{1}$, Guillermo Gonz\'alez-Casado$^{2}$, and
Eduard Salvador-Sol\'e$^{1}$}

\affil{$^{1}$Departamento de Astronom\'\i a y Meteorolog\'\i a, Universidad de
Barcelona, Av.~Diagonal 647, 08028 Barcelona, Spain}
\authoremail{araig@mizar.am.ub.es, eduard@faess0.am.ub.es}

\affil{$^{2}$Departamento  de    Matem\'atica    Aplicada  II, Universidad
  Polit\'ecnica de Catalu\~na, Pau  Gargallo 5, 08028 Barcelona, Spain}
\authoremail{guille@faess0.am.ub.es}

\received{}
\accepted{}

\begin{abstract}
Dark-matter halos show a universal density profile with a scaling such
that less massive systems are typically denser. This mass-density
relation is well described by a proportionality between the
characteristic density of halos and the mean cosmic density at halo
formation time. It has recently been shown that this proportionality could
be the result of the following simple evolutionary picture. Halos form
in major mergers with essentially the same, cosmogony-dependent,
dimensionless profile, and then grow inside-outside, as a consequence
of accretion. Here we verify the consistency of this picture and show
that it predicts the correct zero point of the mass-density relation.
\end{abstract}


\section{INTRODUCTION}

High resolution cosmological N-body simulations show that the
spherically averaged density profile of dark-matter halos has the
following universal form (Navarro, Frenk, \& White 1997, hereafter NFW;
but Moore et al. 1997)
\begin{equation}
{\rho(\xi)\over{\rho_{\rm crit}}}={\delta_{\rm c} \over
\xi(1+\xi)^2}\, .
\label{rho}           
\end{equation}
with $\xi$ the radial distance to the halo center in units of the scale
radius $\rs$, and $\delc$ the characteristic halo density $\roc$ in
units of the critical density for closure $\rho_{\rm crit}$. The
parameters $\rs$ and $\roc$ are linked by the condition that the mean
density within the virial radius $R$ of a halo of a given mass is a
constant factor $a$ times the cosmic critical density; here we take
$a=200$. Therefore, the density profiles of halos at a given epoch
depend on their mass $M$ through one unique parameter, $\delc$ or
$\xs=\rs/R$.

The universality of this profile is not surprising as far as the
gravitational clustering of dark-matter is self-similar in scale-free
cosmogonies and approximately so, over a certain range of masses and
times, in other cosmogonies. However, the specific form of this profile
and the scaling functions $\delc(M)$ or $\xs(M)$ at a given time $t$ is
hard to predict. Much of the research on this issue has focused on the
effects of spherical infall (Gunn \& Gott 1972; Fillmore \& Goldreich
1984; Bertschinger 1985; White \& Zaritsky 1992; Avila-Reese, Firmani,
\& Hern\'andez 1997; Henriksen \& Widrow 1998) and/or secular
evolution owing to dynamical friction and tidal disruption on small
mass clumps (Syer \& White 1997; Nusser \& Sheth 1998). Both these
mechanisms require a certain stability of the system and, therefore,
presume the growth of halos governed by minor mergers or accretion. The
opposite viewpoint that the typical halo density profile is the result
of repetitive, asymmetrical, major mergers producing strong violent
relaxation has also been investigated (e.g., Duncan, Farouki \& Shapiro
1983). In fact, accretion and major mergers alternate in hierarchical
clustering (see, e.g., Tormen, Bouchet \& White 1997 for the CDM
cosmogony). Hence, one should in principle take into account the
combined action of these two extreme processes.

NFW have found that the smaller the mass of halos, the denser they
are. This mass-density correlation was interpreted as reflecting the
fact that, in hierarchical clustering, less massive halos form earlier
when the mean density of the universe is higher. NFW showed, indeed,
that the characteristic density of halos with a given mass at a given
epoch is proportional to the mean cosmic density at their time of
formation. There is, however, a small caveat in this reasoning.
According to it the structural properties of halos would be fixed on
formation. As major mergers yield a substantial rearrangement of the
system, while accretion only causes a smooth secular evolution, it
seems consistent to define the formation of halos as the last major
merger they experience, their identity being kept during
accretion. Yet, the halo formation time used by NFW, drawn from the
extended Press-Schechter (1974, PS) formalism, does not match this
definition.

Recently, Salvador-Sol\'e, Solanes, \& Manrique (1998, SSM) have
repeated the same analysis as NFW but using a better suited formation
time estimate drawn from a modified version of the PS theory. This
incorporates a phenomenological distinction between accretion and major
mergers. For a fractional mass capture threshold for major mergers,
$\Delta_{\rm m}$, equal to 0.6, SSM confirm that, in any hierarchical
cosmogony, there is a clear proportionality between the characteristic
density of halos and the mean density (or still the critical density)
of the universe when they formed. Furthermore, SSM have shown that this
proportionality is consistent with the following simple evolutionary
picture for the structure of dark-matter halos. Halos form, as the
result of major mergers, with essentially universal (i.e., independent
of mass and time), cosmogony-dependent, scaling parameters $\delc$ and
$\xs$. Then, until the next major merger, the corresponding dimensional
parameters $\roc$ and $\rs$ remain fixed, there only being a shift in
$\delc$ and $\xs$ as the critical density decreases and the virial
radius expands accordingly.

The requirement of steady profiles during the accretion phase which
follows from the empirical scaling and the PS clustering model agrees
with accurate dynamical studies of self-similar infall (Fillmore \&
Goldreich 1984; Bertschinger 1985; Henriksen, \& Widrow 1998). This
gives strong support to the validity of that evolutionary
picture. However, its consistency is still to be checked. On the other
hand, although this picture might explain the observed proportionality
between the mass and characteristic density of halos, the zero point of
the mass-density relation in any given cosmogony remains to be
explained. In the present {\it Letter\/} we address these two issues.

\section{THE PREDICTED STRUCTURE OF HALOS FORMED IN MAJOR MERGERS}

In this section we derive, in an arbitrary hierarchical cosmogony, the
values of the scaling parameters $\delc$ or $\xs$ of halos that result,
according to the proposed evolutionary picture, from the major merger
of their typical progenitors with appropriate structural properties.
All elements we need for this calculation are given in SSM except for
the typical mass (and number) of progenitors of a halo with mass $M_0$
at $t_0$. This is next derived in the same frame of the modified PS
clustering model used in SSM.

The specific capture rate, $r^{\rm c}(M'\leftarrow M,t)$, is the rate
at which halos of final mass $M'$ capture, at $t$, other halos of mass
$M<M'$. Following Manrique \& Salvador-Sol\'e (1996) this is related to
the specific major merger rate, $r^{\rm m}(M\rightarrow M',t)$, given
in SSM through
\begin{equation}
r^{\rm c}(M'\leftarrow M,t)=r^{\rm m}(M\rightarrow M',t){N(M,t)\over
N(M',t)} \, ,
\label{scr}
\end{equation}
where $N(M,t)$ is the mass function. For halos with mass $M$ larger
than $M'/(\Delta_{\rm m}+1)$ the specific capture rate is null. This
reflects the fact that such halos evolve by accretion and are,
therefore, identified with the final ones with mass $M'$ which cannot
capture themselves. On the other hand, the capture of halos with masses
$M$ smaller than $M'\delm/ (\Delta_{\rm m}+1)$ do not cause, in
general, the destruction of the capturing halos and, hence, do not
imply the formation of new ones. Only those captures of halos with
masses $M$ ranging between $M'\delm/ (\Delta_{\rm m}+1)$ and
$M'/(\Delta_{\rm m}+1)$ necessarily correspond to major mergers
implying the formation of new halos. Hence, were all major mergers
binary, the specific capture rate given by equation (\ref{scr}) would
be, within that range, a symmetrical function of $M$ around $M'/2$,
which would lead to the equality between the two integrals
\begin{equation} 
C(M',t)\equiv\int_{M'\over 2}^{M'\over (\Delta_{\rm
m}+1)} r^{\rm c}(M'\leftarrow M,t)\,\der M = \int_{M'\delm\over  
(\Delta_{\rm m}+1)}^{M'\over 2}
r^{\rm c}(M'\leftarrow M,t)\,\der M. \label{frbin}
\end{equation}
Moreover, the probability that one of the two progenitors of a halo of
mass $M'$ has mass $M$ within the limits of any of the two preceding
integrals would then be given by
\begin{equation}
\Phi_{\rm p}(M)={r^{\rm c}[M'\leftarrow M,\tf(M',t)]\over 
C(M',t) }\, ,
\label{mpro}
\end{equation}
and the typical masses $M_1$ and $M_2$ of the progenitors of a halo of
mass $M'$ would coincide with the median value of this distribution
function and its complementary value, respectively. N-body simulations
show that major mergers are binary in agreement with the fact that, in
all cosmogonies analyzed, the capture rate (\ref{scr}) is found to be
very closely symmetrical around $M'/2$ in the appropriate range of
masses, the fractional difference between the two integrals in equation
(\ref{frbin}) being less than 3\% at any redshift. Thus, the use of the
probability (\ref{mpro}) to derive the typical mass of progenitors is
fully justified.

Let us consider now two gravitationally bound halo progenitors with
masses $M_1$ and $M_2$ at turnaround. Each mass $M_i$ includes not only
the mass of the main relaxed body, but also that of the surrounding
matter that will have accreted onto it by the time the merger
occurs. Assuming radial orbits, the total energy of the system at
turnaround is
\begin{equation}
E_{\rm ta}=U_1+U_2+E_{12}\, ,
\label{ein}
\end{equation}
with $U_i$ the internal energy of each component, and $E_{12}$ the
interaction energy. If we apply energy conservation to each progenitor
considered in isolation, the internal energy $U_i$ is approximately
equal to the internal energy of the corresponding fully relaxed halo
prior to merger. Assuming spherical symmetry and virial equilibrium,
this is given by
\begin{equation}
U_i=-{1\over 2}{G M_i^2\over R_i}F(\xsi)\, ,
\label{ehalo}
\end{equation}
with $R_i$ and $\xsi$ the virial and dimensionless scale radii,
respectively, of progenitor $i$ at merger time and $F(\xs)$ a
dimensionless function of order unity dependent on the density profile
(\ref{rho}). The interaction energy is also {\it approximately\/} given
by the expression
\begin{equation}
E_{12}=-{G M_1 M_2\over D_{\rm m}}\, ,
\label{erad}
\end{equation}
with $D_{\rm m}$ the turnaround separation between the two centers of
mass, equal to
\begin{equation}
D_{\rm m}^3 = 2GM\left[{\tf(M_0,t_0)\over\pi}\right]^2
\label{din}
\end{equation}
for two point masses falling radially towards each other, where we have
taken into account that the cosmic time corresponding to zero
separation marking the merger of the two progenitors is the formation
time, $\tf(M_0,t_0)$, of the final halo with mass $M_0$ at
$t_0$. Finally, taking into account that the internal energy $U$ of the
final halo at formation, of the form given in equation (\ref{ehalo}),
coincides with the total energy of the system at turnaround
(eq.~\lbrack\ref{ein}\rbrack), we arrive to the approximate relation
\begin{equation}
F(\xsf)\simeq \left({M_1\over M}\right)^{5\over 3} F(x_{\rm s1}) 
+ \left({M_2\over M}\right)^{5\over 3} F(x_{\rm s2}) +
\left({\pi\over 5 \tau}\right)^{2\over 3}
{M_1 M_2\over M^2}\, ,
\label{solve}
\end{equation}
with $M=M_1+M_2$, $\tau$ equal to $\tf(M_0,t_0)$ in units of the Hubble
time at that epoch, and $\xsf$ the value of $\xs$ at halo formation.

In deriving the previous relation we have neglected the external
gravitational torques acting on the system as well as the mutual tidal
interaction between the progenitors. External torques make the
orbits be non-radial which, for a fixed turnaround separation,
increases the infall time. To follow non-radial orbits it is
convenient to use the dimensionless variable $S=J^2 / J_{\rm c}^2(E)$,
where $J$ is the specific angular momentum of $M_2$ relative to $M_1$,
$E$ the corresponding orbital energy, and $J_{\rm c}^2(E)$ the specific
angular momentum of a circular orbit of energy $E$ relative to
$M_1$. For an unperturbed elliptical orbit of eccentricity $e$ one has
$S=1-e$. So the values of $S$ range from $0$ (radial orbit) to $1$
(circular orbit). Taking the merging time equal to one orbital period
leads to
\begin{equation}
E_{12}=\left[{S\over 2(1+M_2/M_1)}-1\right]{G M_1 M_2\over D_{\rm m}}\, ,
\label{e12a}
\end{equation}
\begin{equation}
D_{\rm m}^3=(2-S)^3 G M \left[{\tf(M_0,t_0)\over 2\pi}\right]^2\, .
\label{dina}
\end{equation}
instead of equations (\ref{erad}) and (\ref{din}), respectively. This
introduces an extra factor in the last term on the right-hand of
equation (\ref{solve}) which depends on the non-null value of $S$ and
on $M_2/M_1$. Note that, to be more accurate, we should adopt a
somewhat increased effective value of $S$ to account for the fact that
the merging time for low eccentricity orbits is typically larger than
one single orbital period. Concerning the tidal interaction between the
two progenitors, this causes the so-called tidal braking (part of the
interaction force is spent in deforming the systems which slows down
the orbital motion and increases the infall time) which can also be
accounted for by further increasing the value of $S$. Note that the
importance of external torques is expected to increase with decreasing
mass while, the more massive the halos, the more marked their mutual
tides. Therefore, both effects can be taken into account by adopting,
in a first approximation, a constant positive value of $S$ in the
whole range of masses.

{}From the clustering model given in SSM we can compute the values
of $\tau$, $M_1$, and $M_2$, as well as the typical formation time of
each progenitor. Then, the evolutionary picture for the structure of
halos we wish to check allows us to calculate the values of $\xsi$ of
the two progenitors prior to merger from the assumed universal value of
this parameter on formation. (This can be done by using any of the two
independent fitting formulae given in SSM, the two results typically
differing by less than $10\%$.) Therefore, equation (\ref{solve}) can
be solved for $\xsf$.

\section{COMPARISON WITH THE EMPIRICAL STRUCTURE OF HALOS}

In Figure \ref{snz} we plot, for the NFW density law, the fractional
difference between the solution of equation (\ref{solve}), $\xsf^s$,
and the empirical value of $\xsf$ obtained by SSM used as the formation
value for the progenitors, $\xsf^e$, as a function of $M_0$ at the
present time. $M_0$ is scaled to $M_\star$ defining a unity rms density
fluctuation so that similar graphs hold at other epochs. The curves
plotted here correspond to the standard CDM cosmogony with $b=1.59$
although any other cosmogony analyzed (in particular those studied in
SSM) lead to very similar results. For $S=0$, there is very good
agreement between $\xsf^s$ and $\xsf^e$ at the large mass end. The
deviation increases towards small $M_0$ but, within the four decades
mass range shown in this Figure, it stays below $25\%$. Increasing the
value of $S$ tends to further improve the agreement at small masses
without spoiling too much the good predictions at the large mass end.

Moreover, the evolutionary picture analyzed implies one single,
cosmogony-dependent, value of $\xsf$ fixing the zero point of the
mass-density relation. In Table 1 we quote, for various cosmogonies
(with parameters listed in the first four columns), the value of
$\xsf^e$ minimizing, for $S=0.5$, the maximal fractional difference
between $\xsf^s$ and $\xsf^e$ in the four decades mass range analyzed
(column 5), and the empirical value of $\xsf^e$ obtained in SSM
from the $\chi$-square fit of the mass-density relation (column 6) with
the corresponding $90\%$ confidence interval (column 7). As can be
seen, the theoretical values favored by the evolutionary picture
proposed fall inside the $90\%$ confidence interval associated with the
empirical values in 6 of 8 cases. Only in the case of the flat,
$n=-1.5$, and the open, $n=-1$, cosmogonies the theoretical prediction
in column (5) is a little too small. In fact, the trend for the
theoretical value to be slightly smaller than the empirical one seems
quite general suggesting the existence of some slight bias of the kind
mentioned in SSM. In any event, the general agreement between the
theoretical and empirical values of $\xsf$ in all cosmogonies
considered is hardly a coincidence. This strongly supports the validity
of the schematic evolutionary picture for dark-matter halos proposed in
SSM.

\section{DISCUSSION}

In \S\ 2 we have considered only elliptical orbits with different
possible values of $S$ to account for their unknown typical
eccentricity. These correspond, indeed, to the most reasonable initial
condition for mergers regardless the cosmogony. Mergers from parabolic
or hyperbolic orbits are certainly much more improbable because of
their non-negative orbital energy. This seems to contradict the idea
that in scale free cosmogonies (i.e., an Einstein-de Sitter universe
with power law spectrum: $P(k)\propto k^n$) with $n=1$, mergers occur
typically between halos describing parabolic orbits, while for $n>1$
($n<1$) they would follow hyperbolic (elliptical) orbits. The reason is
that the mean cosmic density and, hence, the mean density of halos with
characteristic mass $M_*$, scales as $M_*^{-(3+n)/2}$. Therefore, for
$n=1$ one has $M_*\propto R$ implying that the specific binding energy
of such halos is constant in time. One must be very cautious with this
kind of qualitative arguments. In the previous reasoning, the binding
energy refers only to a very specific class of halos, those which have
the characteristic mass at the given epoch. Furthermore, the typical
progenitors of such halos will hardly coincide with the class of
objects which have characteristic mass at the adequate epoch.

In this {\it Letter\/} we have checked the consistency, relative to
energy conservation, of the scaling evolution inferred in SSM from the
empirical mass-density relation. We must stress however that such a
relation is not the simple consequence of energy conservation in
mergers. Our results show that the mass-density relation depends, in
addition, on the assumed evolution of halo structure. Any departure
from the concrete evolutionary picture proposed (e.g., other values of
parameters $\delc$ and $\xs$ at halo formation or a non-steady behavior
of halos during accretion) would have led, in general, to a mass-density
relation in disagreement with $N$-body simulations.

\acknowledgments This work has been supported by the Direcci\'on
General de Investigaci\'on Cient\'\i fica y T\'ecnica under contract
PB96-0173.

\clearpage

\figcaption[fig1]{Fractional difference between the value on formation
of the scaling parameter $\xs$ obtained for halos of mass $M_0$ and the
one assumed for their corresponding progenitors ($\xsf^s$ and
$\xsf^e$, respectively), for the case of the NFW universal density law
and the standard CDM cosmogony. The values of the orbital parameter $S$
used range from zero (upper curve) to one (lower curve) in steps of
0.25. Solid lines correspond to $\xsf^e$ equal to the empirical value
of $\xs$ found in SSM. In dashed line, the best theoretical prediction
obtained (for any fixed value of $S$) by taking $\xsf^e$ as a free
parameter.
\label{snz}}

\clearpage

\begin{deluxetable}{lrrrccccc}
\tablewidth{0pt}
\tablenum{1}
\tablecaption{Scaling of Halo Density Profiles\label{param}}
\tablehead{
\colhead{$P(k)$}                  & \colhead{$\Omega_0$}       &   
\colhead{$\lambda_0$}             & \colhead{$\sigma_8$}       &
\colhead{$x_{\rm sf}$\tablenotemark{a}} &
\colhead{$x_{\rm sf}$\tablenotemark{b}}  & 
\colhead{$90\%$ c.i.}   
\nl \tablevspace{5pt}
\colhead{(1)}                  & \colhead{(2)}            &
\colhead{(3)}                  & \colhead{(4)}            &
\colhead{(5)}                  & \colhead{(6)}            &
\colhead{(7)}}
\tablenotetext{a}{best theoretical prediction (for S=0.5)}
\tablenotetext{b}{best fit to the mass-density relation}
\startdata
SCDM & 1.0 & 0.0 & 0.63 & 0.14 & 0.17 & (0.12,0.25) \nl
$\Lambda$CDM & .25 & .75 & 1.3 & 0.18 & 0.29 & (0.18,0.52)\nl
$n=-1.5$ & 1.0 & 0.0 & 1.0 & 0.09 & 0.20 & (0.12,0.36) \nl
$n=-1.0$ & 1.0 & 0.0 & 1.0 & 0.10 & 0.17 & (0.10,0.30) \nl
	& 0.1 & 0.0 & 1.0 & 0.08 & 0.19 & (0.12,0.31) \nl
$n=-0.5$ & 1.0 & 0.0 & 1.0 & 0.10 & 0.14 & (0.08,0.23) \nl
$n= 0.0$ & 1.0 & 0.0 & 1.0 & 0.09 & 0.09 & (0.06,0.14) \nl
        & 0.1 & 0.0 & 1.0 & 0.10 & 0.06 & (0.04,0.12) \nl
\enddata
\end{deluxetable}

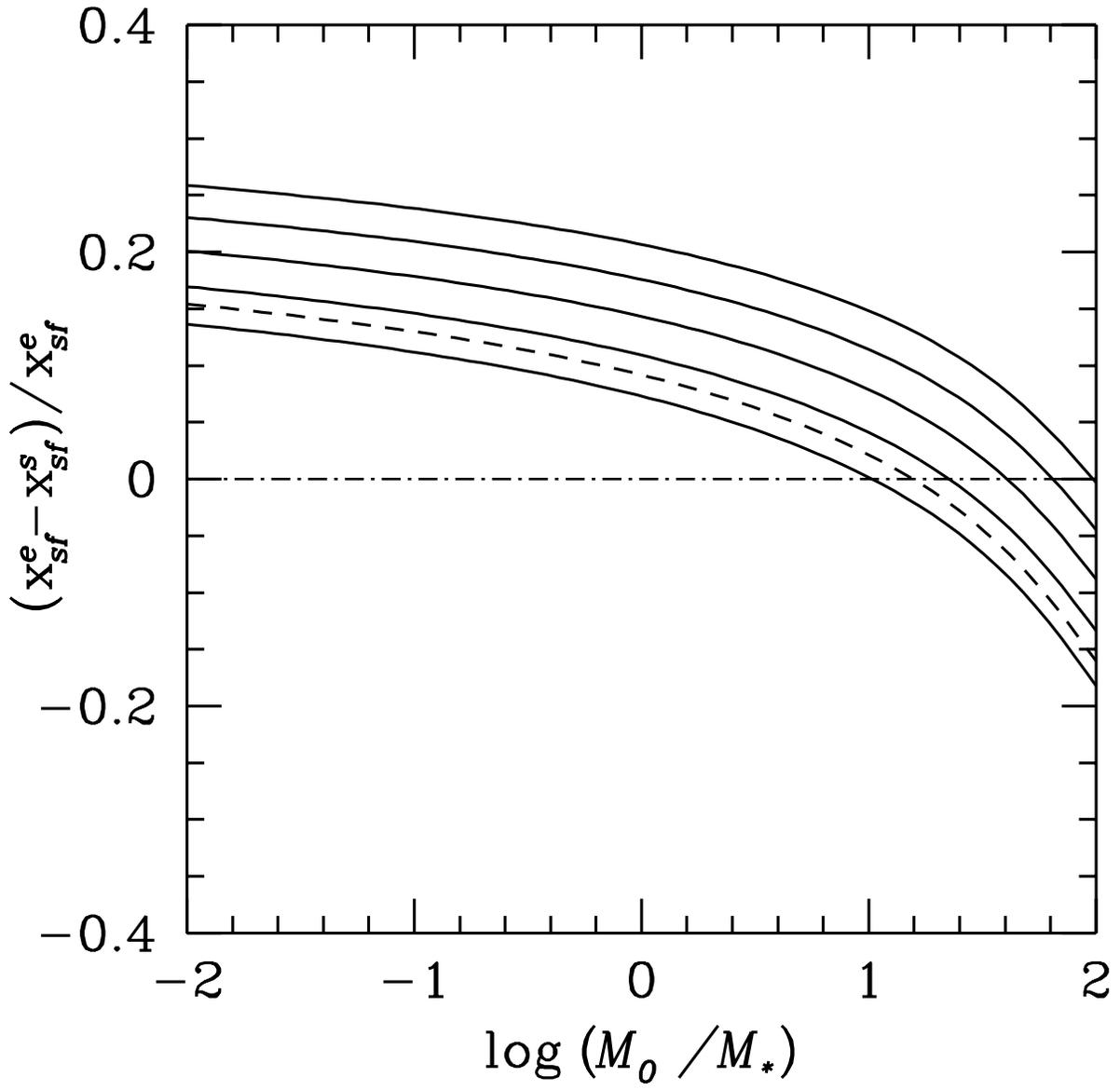
\end{document}